\documentclass[11pt,a4paper]{article}
\pdfoutput=1
\usepackage{jheppub}

\usepackage[]{amsmath}
	\numberwithin{equation}{section}
\usepackage[]{amssymb}
\usepackage{amsfonts}

\DeclareMathOperator{\R}{\mathbb{R}}
\DeclareMathOperator{\Zint}{\mathbb{Z}}

\DeclareMathOperator{\cs}{\mathbb{S}}
\newcommand{\I}{\mathrm{i}}

\newcommand{\fre}{\mathcal{F}}

\usepackage{hyperref}

\usepackage[toc,page]{appendix}

\usepackage[utf8]{inputenc}
\usepackage[english]{babel}

\usepackage{enumerate}

\usepackage{graphicx}

\title{Large $N$ phase transition in $T\overline{T}$-deformed $2d$ Yang--Mills theory on the sphere}
\author{Leonardo Santilli}
\author{and Miguel Tierz}
\affiliation{Departamento de Matem\'{a}tica, Grupo de F\'{\i}sica Matem\'{a}tica, Faculdade de Ci\^{e}ncias, Universidade de Lisboa, Campo Grande, Edif\'{\i}cio C6, 1749-016 Lisboa, Portugal.}
\emailAdd{lsantilli@fc.ul.pt}
\emailAdd{tierz@fc.ul.pt}
\abstract{We study the partition function of a $T \overline{T}$-deformed version of Yang--Mills theory on the two-sphere. We show that the Douglas--Kazakov 
phase transition persists for a range of values of the deformation parameter, and that the critical area is lowered. The transition is of third order and also induced by instantons, whose contributions we characterize.}

\begin{document}

	\maketitle

\section{Introduction}

Low dimensional quantum field theories have been proved for decades to be a
very valuable source of exact results, providing numerous insights into
quantum theory as well as showing many direct relationships with statistical
mechanical systems, strongly correlated systems and integrable systems, just
to name a few. In recent years, a certain deformation of two-dimensional
relativistic quantum field theories, based on the specific properties,
described by Zamolodchikov in \cite{Zamolodchikov:2004ce}, of the $T_{zz}T_{%
\bar{z}\bar{z}}-T_{z\bar{z}}^{2}$ operator, also known as $T\overline{T}$
operator and where $T_{zz}$ denotes the components of the stress tensor in
complex coordinates, is attracting a considerable amount of interest.

The vacuum expectation of the operator has distinctive properties \cite%
{Zamolodchikov:2004ce} and irrelevant deformations by this operator were
studied in \cite{Smirnov:2016lqw,Cavaglia:2016oda}, showing that, after
compactification of the theory on a Euclidean circle of radius $R$, a simple
differential equation (a Burgers equation) governs how the energy spectrum
at finite $R$ evolves according to a parameter $\tau$, which controls the
strength of the deformation by the $T\overline{T}$ operator.

A large number of works involving this deformation have appeared already,
including work showing that the deformation is equivalent to coupling the
theory to flat space Jackiw--Teitelboim gravity such that at high energies
is a gravitational theory with no local degrees of freedom \cite%
{Dubovsky:2017cnj,Dubovsky:2018bmo} and a number of other results (see for
example \cite%
{Giveon:2017nie,Giveon:2017myj,Datta:2018thy,Aharony:2018bad,Cardy:2018sdv,Conti:2018tca}%
), including also results for non-relativistic systems \cite{Cardy:2018jho}.

We will not be using any of these developments specifically but, rather,
will take on the last part of the recent work \cite{Conti:2018jho}, where
the $T\overline{T}$-deformation of two-dimensional Yang--Mills theory \cite%
{Cordes:1994fc} is presented. We shall focus on the case of the sphere and
see how two of the most salient features of the original theory still hold,
but are modified by the deformation. Namely, the Douglas--Kazakov large $N$
phase transition of the theory and the proof that such phase transition is
induced by (unstable) instantons.

The paper is organized as follows: we introduce below the basics of
two-dimensional Yang--Mills theory and its $T\overline{T}$-deformation. In Section \ref{sec:TTlargeN}, we study the large $N$ behavior of the partition function,
following \cite{Douglas:1993iia}, and obtain a phase transition with a
critical area modified by a factor, that we fully specify. We characterize
both the weak-coupling and the strong-coupling phase, with the former still
being described by a Wigner semicircle distribution for the eigenvalues, but
with a nontrivial rescaling of the area parameter. Note that the existence
of the weak-coupling phase is dependent on the value of the $\tau $
parameter. In particular, there are values of the
deformation, given by $\tau \geq 12\pi ^{2}/(12-\pi ^{2})$, for
which there is only the strong-coupling phase.

As happens in the undeformed case, the recursion involved in the
strong-coupling phase is more complicated, but we characterize the
nonperturbative solution near the critical point. The phase transition is
found to be of third order, as in the undeformed case.

In Section \ref{sec:instantons}, we explore what happens with the well-known
and elegant explanation of the phase transition as triggered by instantons 
\cite{Gross:1994mr}. For this, we use the corresponding matrix model
description for the deformed theory and obtain that the same mechanism is at
work, and moreover we are able to quantify the (stronger) effects of the
instantonic contributions. Finally, we conclude with avenues for further
work.

\subsection{Two-dimensional Yang--Mills theory and its $T\overline{T}$%
-deformation}

We review quantum Yang--Mills theory with gauge group $SU(N)$ on an oriented
closed Riemann surface $\Sigma _{h}$ of genus $h$ and unit area form $%
\mathrm{d}\mu $~\cite{Cordes:1994fc}. The action is%
\begin{equation}
S_{\mathrm{YM}}=-\frac{1}{4g^{2}}\,\int_{\Sigma _{h}}\,\mathrm{d}\mu ~%
\mathrm{Tr}\,F^{2}\ ,  \label{continuum}
\end{equation}%
where $g^{2}$ plays the role of the coupling constant, $F$ is the field
strength of a matrix gauge connection, and $\mathrm{Tr}$ is the trace in the
fundamental representation of $SU(N)$. Migdal discussed the lattice
regularization of the gauge theory, which relies on a triangulation of the
two-dimensional manifold $\Sigma _{h}$ with group matrices situated along
the edges~\cite{Migdal:1975zg}. The path integral is then approximated by
the finite-dimensional unitary matrix integral 
\begin{equation}
\mathcal{Z}_{\mathrm{M}}=\int \,\prod_{\mathrm{edges}~\ell }\,\mathrm{d}%
U_{\ell }~\prod\limits_{\mathrm{plaquettes}~P}\,Z_{P}\left[ U_{P}\right] \ ,
\label{latticeint}
\end{equation}%
where $\mathrm{d}U_{\ell }$ denotes Haar measure on $SU(N)$ and the holonomy 
$U_{P}=\prod\nolimits_{\ell \in P}\,U_{\ell }$ is the ordered product of
group matrices along the links of a given plaquette. The local factor $Z_{P}%
\left[ U_{P}\right] $ is a suitable gauge invariant lattice weight that
converges in the continuum limit to the Boltzmann weight for the Yang--Mills
action \eqref{continuum}. The use of the heat kernel lattice action for the
lattice weight $Z_{P}\left[ U_{P}\right] $, has very relevant features \cite%
{Menotti:1981ry} and is the usual choice in two-dimensional Yang--Mills
theory~\cite{Cordes:1994fc}. It leads to the group theory expansion of the
partition function~\cite{Migdal:1975zg,Rusakov:1990rs}%
\begin{equation}
\mathcal{Z}_{\mathrm{M}}=\sum_{R}\,\left( \dim R\right) ^{2-2h}\,\exp \big(-%
\frac{g^{2}A}{2}\,C_{2}(R)\big)\ ,  \label{HK}
\end{equation}%
where the sum runs over all isomorphism classes $R$ of irreducible
representations of the $SU(N)$ gauge group, $\dim R$ is the dimension of the
representation $R$ and $C_{2}(R)$ is the quadratic Casimir invariant of $R$.
The $g^{2}$ is the Yang--Mills coupling, and $A$ is a parameter which can be
identified with the area of the surface. In the following, we shall focus on 
$U(N)$ gauge theory, instead of $SU(N)$, so we use the identification $%
U(N)=SU(N)\times U(1)/\Zint_{N}$, and the heat kernel expansion \eqref{HK}
remains the same but with the sum now running over classes of irreducible
representations of $U(N)$. In terms of the partition $(r_{1},\dots ,r_{N})$
associated to the irreducible representation $R$, with 
\begin{equation}
r_{1}\geq r_{2}\geq \dots \geq r_{N},  \label{eq:cond1}
\end{equation}%
the dimension of the representation is given by Weyl's denominator formula%
\begin{equation}
\dim R=\prod_{1\leq i<j\leq N}\frac{r_{i}-r_{j}+j-i}{j-i},
\end{equation}%
and the Casimir is 
\begin{equation}
C_{2}\left( R\right) =\sum_{i=1}^{N}r_{i}\left( r_{i}-2i+N+1\right) .
\end{equation}%
The quantities depending on representations of $U(N)$ can also be obtained
using the correspondence of two-dimensional Yang--Mills with quantum
mechanics on the group manifold $U(N)$ \cite{Douglas:1993wy}. Throughout
this work, we will focus on the case of the two-sphere, corresponding to $%
h=0 $ in the formula \eqref{HK} above.

The large $N$ limit of \eqref{HK} was studied in \cite%
{Douglas:1993iia,Minahan:1993tp,Gross:1994mr}, showing the presence of a
phase transition, known as Douglas--Kazakov (DK) phase transition \cite%
{Douglas:1993iia}. As was the case with the Gross--Witten--Wadia phase
transition \cite{Gross:1980he,Wadia:1980cp,Wadia:2012fr}, it is of third order. In \cite%
{Gross:1994mr} it was shown how, from the point of view of the small
coupling phase, the transition is triggered by instantons. New light on this
phenomenon was shed in \cite{Caselle:1993mq,Minahan:1993tp,Douglas:1993wy},
where Yang--Mills theory on the two-sphere was reformulated as a
nonrelativistic fermionic system, and from the interpretation of the theory
as a string theory, by expressing the partition function as a weighted sum
over maps from a two-dimensional worldsheet into the manifold \cite%
{Gross:1992tu,Gross:1993hu} (see also the review \cite{Cordes:1994fc} and
the recent \cite{Donnelly:2016jet}).

In this work we analyze the large $N$ behaviour of $T \overline{T}$-deformed Yang--Mills theory on $\cs^{2}$. In \cite{Conti:2018jho}, a $T \overline{T}$-deformation of the $2d$ Yang--Mills Lagrangian and Hamiltonian has been obtained. For the specific case of $2d$ Yang--Mills theory, the use of the Hamiltonian formalism is more convenient. As mentioned above, the energy levels of the deformed theory are known to evolve with the strength of the deformation, following a differential equation of Burgers type \cite{Smirnov:2016lqw,Cavaglia:2016oda}. Passing from gauge fields to their conjugate momenta, only one variable is non-vanishing: as a consequence, the differential equation satisfied by the $2d$ Yang--Mills Hamiltonian $\mathcal{H}_{\mathrm{YM}}$ reduces to the one of a ``pure potential'' term. The $T \overline{T}$-deformation in this case takes a specially simple form \cite{Conti:2018jho}:
\begin{equation*}
	\mathcal{H}_{\mathrm{YM}} \mapsto \frac{ \mathcal{H}_{\mathrm{YM}} }{1- \tau \mathcal{H}_{\mathrm{YM}} } .
\end{equation*}
The Hamiltonian of Yang--Mills theory on a closed oriented Riemann surface $\Sigma_h$ is diagonal in the representation basis, with eigenvalues given by the quadratic Casimir of the irreducible representations \cite{Cordes:1994fc}. Therefore the above deformation as a pure potential becomes\footnote{Notice that, with respect to \cite{Conti:2018jho}, we scaled the parameter $\tau $ by a factor $\frac{2}{N^{2}}$ and set $\lambda = g^2 N \equiv 1$.}:
\begin{equation}
\label{eq:TTdefC2R}
C_{2}\left( R\right) \longmapsto \frac{C_{2}\left( R\right) }{1-\frac{\tau }{N^{3}}C_{2}\left( R\right) } ,
\end{equation}
and the theory is described by the partition function \eqref{HK} with the quadratic Casimir replaced by \eqref{eq:TTdefC2R}.

\section{Large $N$ limit of the $T \overline{T}$-deformed theory}
\label{sec:TTlargeN}

We study the partition function of $T \overline{T}$-deformed $U(N)$
Yang--Mills on $\cs^2$: 
\begin{equation}  \label{eq:ttZYM2}
\mathcal{Z} _N (A, \tau) = \sum_{R} \left( \dim R \right)^2 \exp \left( - 
\frac{g^2 A}{2} \left( \frac{C_2 \left( R \right)}{1 - \frac{ \tau }{N^3}
C_2 \left( R \right) } \right) \right) .
\end{equation}
As customary, we introduce the `t Hooft parameter 
\begin{equation*}
\lambda \equiv g^2 N
\end{equation*}
to be held fixed at large $N$. As the partition function only depends on the
combination $\lambda A$, we define $A^{\prime} := \lambda A$ and drop the
prime from now on, hence identifying the area $A$ with the coupling.

Following the standard procedure for the large $N$ analysis \cite%
{Douglas:1993iia}, we introduce the variables 
\begin{equation}
x:=\frac{i}{N},\quad r(x):=\frac{r_{i}}{N},\quad h(x):=-r(x)+x-\frac{1}{2},
\end{equation}%
and, sending $N\rightarrow \infty $, replace $N^{-1}\sum_{i=1}^{N}$ with $%
\int_{0}^{1}dx$. The dimension of a representation $R$ therefore becomes 
\begin{equation}
\dim R=\exp \left\{ N^{2}\left[ \int_{0}^{1}dx\int_{x}^{1}dy\log \lvert
h(y)-h(x)\rvert -\log \lvert y-x\rvert \right] \right\}
\end{equation}%
while the quadratic Casimir reads 
\begin{equation}
\frac{1}{N}C_{2}\left( R\right) =N^{2}\int_{0}^{1}dx\left( h(x)-\left( x-%
\frac{1}{2}\right) \right) \left( h(x)+\left( x-\frac{1}{2}\right) \right) .
\end{equation}%
In the same way, condition \eqref{eq:cond1} becomes 
\begin{equation}
\frac{h(y)-h(x)}{y-x}\geq 1.  \label{eq:cond2}
\end{equation}%
At this point, we assume values of $(A,\tau) $ such that $\frac{\tau }{N^{3}}%
C_{2}(R)<1$ for every $R$. As we will see later, this is not inconsistent
and corresponds to introduce a $\tau $-dependent lower bound in the region
of validity of the solution as a function of $A$, that is, $A>A_{lb}$. Under
this condition, we can expand the function in the exponential of %
\eqref{eq:ttZYM2} as a geometric series.

Putting all together, the large $N$ limit of the partition function %
\eqref{eq:ttZYM2} is: 
\begin{equation}
\mathcal{Z} (A,\tau )=\int \mathcal{D}he^{-N^{2}S\left[ h\right] },
\end{equation}%
where $\mathcal{D}h$ is some measure on the space of functions supported in $%
[0,1]$ satisfying \eqref{eq:cond2}, and 
\begin{equation}
S\left[ h\right] =-\int_{0}^{1}dx\int_{0}^{1}dy\log \lvert h(y)-h(x)\rvert -%
\frac{3}{2}+\frac{A}{2}\sum_{j=0}^{\infty }\tau ^{j}\left[
\int_{0}^{1}dxh(x)^{2}-\frac{1}{12}\right] ^{j+1}.  \label{eq:SofH}
\end{equation}%
At this point, we introduce the eigenvalue density $\rho $, as usual,
according to: 
\begin{equation*}
\rho (h)dh=dx,
\end{equation*}%
which is normalized: 
\begin{equation}
\int dh\rho (h)=1.
\end{equation}%
The action functional \eqref{eq:SofH} becomes: 
\begin{equation}
S\left[ \rho \right] =-\int du\rho (u)\int dv\rho (v)\log \lvert u-v\rvert -%
\frac{3}{2}+\frac{A}{2}\sum_{j=0}^{\infty }\tau ^{j}\left[ \int du\rho
(u)u^{2}-\frac{1}{12}\right] ^{j+1},
\end{equation}%
and the constraint \eqref{eq:cond2} imposes the condition on the eigenvalue
distribution 
\begin{equation}
\rho (h)\leq 1.  \label{eq:condrho}
\end{equation}%
In the infinite $N$ limit, the partition function receives only the
contribution by the distribution $\rho $ which solves the saddle point
equation $\frac{\delta S}{\delta h}=0$. That is, we pursue a distribution $%
\rho $ satisfying the integral equation 
\begin{equation}
-2\mathrm{P}\int du\frac{\rho (u)}{h-u}+Ah\sum_{j=0}^{\infty }\left(
j+1\right) \tau ^{j}\left[ \int du\rho (u)u^{2}-\frac{1}{12}\right] ^{j}=0,
\label{eq:SPE}
\end{equation}%
where the symbol $\mathrm{P}\int $ means the principal value of the integral.

The saddle point equation \eqref{eq:SPE} is hard to solve analytically, due
to the $\rho $ appearing in the geometric series. Nevertheless, the $T%
\overline{T}$-deformation only introduced powers of the second moment of the
eigenvalue distribution, which is a well defined quantity, and the
dependence on $h$ remains factorized. This allows for a perturbative
solution in $\tau $, and we will solve equation \eqref{eq:SPE} to all orders.

\subsection{Perturbative solution}

At zero-th order in $\tau $, the theory obviously reduces to pure
Yang--Mills and equation \eqref{eq:SPE} describes the Douglas--Kazakov
distribution \cite{Douglas:1993iia}. Indeed, for $j=0$ the equation reduces
to the saddle point equation of a Gaussian matrix model: 
\begin{equation}
\mathrm{P}\int du\frac{\rho (u)}{h-u}=\frac{A}{2}h,
\end{equation}%
which is solved by the celebrated Wigner semicircle distribution 
\begin{equation}
\rho (h)=\frac{A}{2\pi }\sqrt{\frac{4}{A}-h^{2}},\qquad \mathrm{supp}\rho =%
\left[ -\frac{2}{\sqrt{A}},\frac{2}{\sqrt{A}}\right] .  \label{eq:rho0sc}
\end{equation}%
However, the solution must satisfy the constraint \eqref{eq:condrho} $\rho
\leq 1$, meaning that the present one-cut solution only holds up to $%
A_{cr}^{(0)}=\pi ^{2}$. For the moment, we focus on the perturbative
analysis in the small coupling phase $A<A_{cr}$, and discuss the strong
coupling phase $A>A_{cr}$ in the next subsection.

The second moment of the Wigner semicircle distribution \eqref{eq:rho0sc}
is: 
\begin{equation}
\frac{A}{2\pi }\int_{-2/\sqrt{A}}^{2/\sqrt{A}}dh\left( \sqrt{\frac{4}{A}%
-h^{2}}\right) h^{2}=\frac{1}{A}.
\end{equation}%
A check of the consistency condition for the geometric expansion at this
order: 
\begin{equation*}
\tau \left( \frac{1}{A}-\frac{1}{12}\right) <1,
\end{equation*}%
leads to 
\begin{equation}
A>A_{lb}^{(0)}=\frac{12\tau }{12+\tau }.
\end{equation}%
In particular, this restriction is removed when $\tau \rightarrow 0$, as it
should for the undeformed limit. We remark, however, that this is an $%
\mathcal{O} (1)$ estimation of $A_{lb}$, and not a true constraint, which
must be imposed on the full (nonperturbative) result.

\medskip We now proceed to the next order in perturbation theory,
corresponding to $j=0,1$ in the geometric expansion. The saddle point
equation at order $\tau $ is: 
\begin{equation}
\mathrm{P}\int du\frac{\rho (u)}{h-u}=\frac{Ab_{1}}{2}h,  \label{eq:SPE_sc1}
\end{equation}%
where we have denoted 
\begin{equation*}
b_{1} \equiv b_{1}\left( A,\tau \right) =1+\tau \left( \frac{1}{A}-\frac{1}{%
12}\right) .
\end{equation*}%
As we are in the small coupling phase, $A<A_{cr}^{(0)}=\pi ^{2}$, we have
that $b_{1}(A,\tau )>1$. The saddle point equation \eqref{eq:SPE_sc1} is
again satisfied by the Wigner semicircular distribution, now with parameter $%
Ab_{1}$, that is: 
\begin{equation}
\rho (h)=\frac{Ab_{1}}{2\pi }\sqrt{\frac{4}{Ab_{1}}-h^{2}},\qquad \mathrm{%
supp}\rho =\left[ -\frac{2}{\sqrt{Ab_{1}}},\frac{2}{\sqrt{Ab_{1}}}\right] .
\end{equation}%
From this it stems that the second moment at first order in $\tau $ is $%
1/Ab_{1}$. The constraint \eqref{eq:condrho} implies $Ab_{1}<\pi ^{2}$ and,
as $b_{1}>1$, in particular we get 
\begin{equation*}
\frac{\pi ^{2}-2\tau }{1-\frac{\tau }{6}}=A_{cr}^{(1)}<A_{cr}^{(0)}=\pi ^{2}.
\end{equation*}

We now consider a generic order $k$ in the perturbative expansion in the
parameter $\tau $. The general procedure is clear from order 1, and can be
iterated, giving order by order a Wigner semicircle distribution with
different coefficients. The second moment, approximated at previous order,
is $1/Ab_{k-1}$, and the saddle point equation reduces to 
\begin{equation}
\mathrm{P}\int du\frac{\rho (u)}{h-u}=\frac{Ab_{k}}{2}h,
\end{equation}%
with generic multiplicative factor 
\begin{equation}
b_{k} \equiv b_{k}(A,\tau )=\sum_{j=0}^{k}\left( j+1\right) t^{j}\left( 
\frac{1}{Ab_{k-1}}-\frac{1}{12}\right) ^{j}.  \label{eq:recursiveexpr}
\end{equation}%
Notice that we have a recursive way to calculate the $b_{k}$'s, only
depending on the previous one, although in a nontrivial way.

The solution is given by 
\begin{equation}
\rho (h) = \frac{A b_k }{2 \pi} \sqrt{ \frac{4}{A b_k} - h^2 } , \qquad 
\mathrm{supp} \rho = \left[ - \frac{2}{\sqrt{A b_k}} , \frac{2}{\sqrt{A b_k}}
\right] ,
\end{equation}
as long as the condition $A b_k < \pi^2$ holds. In particular, as $b_k = 1 + 
\mathcal{O} \left( \tau \right)$, we have that the critical value of the
area is lowered from the pure Yang--Mills case, i.e. $A_{cr} ^{(k)} < \pi^2$%
, at least for $\tau$ small enough. Consistently, the constraint guarantees
order by order that: 
\begin{equation*}
\tau \left( \frac{1}{A b_k } - \frac{1}{12} \right) \ge \tau \left( \frac{1}{%
\pi^2 } - \frac{1}{12} \right) \ge 0.
\end{equation*}

We will now obtain the full solution to \eqref{eq:SPE} by including all
orders in $\tau $. This formally corresponds to evaluate recursive relation %
\eqref{eq:recursiveexpr} for all $k$, and the eigenvalue distribution is
then given by the Wigner semicircle expression with parameter $Ab_{\infty }$%
. From expression \eqref{eq:recursiveexpr} one recursively infers that 
\begin{equation*}
b_{k}\leq b_{k-1}+(k+1)\tau ^{k}\left( \frac{1}{Ab_{k-1}}-\frac{1}{12}%
\right) ^{k}\quad \Longrightarrow \quad |b_{k}-b_{k-1}|\rightarrow 0,
\end{equation*}%
and therefore $b_{\infty }(A,\tau )$ is given by the solution of the
equation: 
\begin{equation}
b_{\infty }=\lim_{k\rightarrow \infty }\sum_{j=0}^{k}\left( j+1\right)
t^{j}\left( \frac{1}{Ab_{k-1}}-\frac{1}{12}\right) ^{j}.  \label{binf}
\end{equation}%
Samples of the convergence of $b_{k}$ are given in Figure \ref%
{fig:bkconvergence}. Writing the right hand side of (\ref{binf}) as the
derivative of a geometric series\footnote{%
We can do that for $A>A_{lb}$, as we assumed at the beginning.}, $b_{\infty
} $ is determined by solving: 
\begin{equation}
b_{\infty }=\left[ 1-\tau \left( \frac{1}{Ab_{\infty }}-\frac{1}{12}\right) %
\right] ^{-2}.  \label{eq:defBinf}
\end{equation}%
\begin{figure}[th]
\centering
\includegraphics[width=0.4\textwidth]{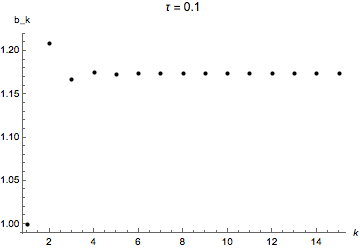}\hspace{0.1\textwidth} %
\includegraphics[width=0.4\textwidth]{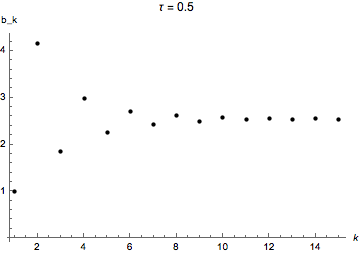}
\caption{Convergence of the sequence $\left\{ b_{k}\right\} _{k}$, for $%
\protect\tau =0.1$ (left) and $\protect\tau =0.5$ (right).}
\label{fig:bkconvergence}
\end{figure}
This leads to a cubic equation in $b_{\infty }$, but only one of the three
solutions satisfies 
\begin{equation*}
b_{\infty }(A,\tau )\xrightarrow{\tau \to 0}1,
\end{equation*}%
hence we uniquely identify our solution for $b_{\infty }(A,\tau )$.
Explicitly: 
\begin{equation}
b_{\infty }(A,\tau )=\frac{1+\frac{2\tau }{A}\left( 1+\frac{\tau }{12}%
\right) +\sqrt{1+\frac{4\tau }{A}\left( 1+\frac{\tau }{12}\right) }}{2\left(
1+\frac{\tau }{12}\right) ^{2}}.  \label{eq:b_infinity}
\end{equation}%

The small coupling region is defined by the condition $Ab_{\infty }<\pi ^{2}$%
, whence the critical value for the coupling is 
\begin{equation}
A_{cr}(\tau )=\pi ^{2}\left( 1-\tau \left( \frac{1}{\pi ^{2}}-\frac{1}{12}%
\right) \right) ^{2} ,
\end{equation}%
as long as $\tau < \frac{12 \pi^2}{12 - \pi^2}$, and no positive solution for $\tau$ bigger than the mentioned value.
At this point, we ought to check the consistency of our initial assumption:
we developed a perturbative expansion in $\tau $, and then solved it to all
orders, assuming the existence of a region $A>A_{lb}$ for which 
\begin{equation*}
\tau \left( \frac{1}{Ab_{\infty }}-\frac{1}{12}\right) <1,
\end{equation*}%
corresponding to: 
\begin{equation}
Ab_{\infty }(A,\tau )>\frac{12\tau }{12+\tau }.  \label{Abinf}
\end{equation}%
As we have the explicit expression for $b_{\infty }(A,\tau )$, we can see
that the infimum of the left hand side of (\ref{Abinf}), as a function of $A$
is exactly the right hand side, that is, $b_{\infty }$ takes exactly the
expression for which the lower bound is pushed to $A_{lb}=0$. This means
that our procedure holds for any $A>0,\tau \geq 0$, or, in other words, the
assumption we made to start with the perturbative procedure is always
verified in the region of validity of the heat kernel expansion.

To summarize, we have proved that, after the $T\overline{T}$-deformation, we
still have a small coupling phase $0<A<A_{cr}(\tau )$ analogous to the
undeformed case, with eigenvalue distribution given by a Wigner semicircle.
Nevertheless, the effect of the deformation is to modify the parameter of
the distribution, as well as moving the original critical value \cite%
{Douglas:1993iia}. In particular, for small values of $\tau $, the value $A_{cr}(\tau
) $ of the critical area is a decreasing function, hence $A_{cr}\leq \pi ^{2}$, whilst when $\tau
\ge \frac{12\pi ^{2}}{12-\pi ^{2}}$ we do not have any
small coupling phase, due to the constraint $A>0$, and only the strong
coupling phase exists\footnote{%
Notice that for $A\leq 0$ the theory is ill-defined even in the undeformed
model. It can easily be seen from expression \eqref{HK}, which is not
convergent for nonpositive values of $A$.}. The eigenvalue density in the
small coupling phase $0<A<A_{cr}(\tau )$ is: 
\begin{equation}
\rho (h)=\frac{Ab_{\infty }(A,\tau )}{2\pi }\sqrt{\frac{4}{Ab_{\infty
}(A,\tau )}-h^{2}},\qquad \mathrm{supp}\rho =\left[ -\frac{2}{\sqrt{%
Ab_{\infty }(A,\tau )}},\frac{2}{\sqrt{Ab_{\infty }(A,\tau )}}\right] ,
\label{eq:FullWignerSmallA}
\end{equation}%
where $b_{\infty }(A,\tau )$ is given in \eqref{eq:b_infinity}.

\subsection{Strong coupling phase}

Throughout the solution showed above, we had to impose an upper bound to the
coupling $A$ in order not to violate the constraint \eqref{eq:condrho}. When 
$A>A_{cr}(\tau )$ the Wigner semicircle distribution is not allowed anymore,
so that we have to look for a two-cut solution of the saddle point equation %
\eqref{eq:SPE}. We follow again a perturbative approach, reproducing the
procedure of \cite{Douglas:1993iia} order by order. According to what we
have seen in the small coupling phase, the $T\overline{T}$-deformation
introduces a nontrivial dependence on the parameters $A$ and $\tau $, but
preserves the form of the Douglas--Kazakov solution. Therefore the two-cut
solution, if any, must be of the form: 
\begin{equation*}
\rho (h)=%
\begin{cases}
\varphi (h),\quad h\in \left[ -\alpha ,-\beta \right] \cup \left[ \beta
,\alpha \right] , \\ 
1,\quad \quad \ \ h\in \left[ -\beta ,\beta \right] ,%
\end{cases}%
\end{equation*}%
where $0\leq \beta \leq \alpha $ depend, in general, on $A$ and $\tau $.
Plugging this expression into \eqref{eq:SPE} we get: 
\begin{equation}
\mathrm{P}\int du\frac{\varphi (u)}{h-u}=\log \left( \frac{h-b}{h+b}\right) +%
\frac{A}{2}h\sum_{j=0}^{\infty }\left( j+1\right) \tau ^{j}\left[ \int
du\rho (u)u^{2}-\frac{1}{12}\right] ^{j}.  \label{eq:SPE2}
\end{equation}%
The idea is again to proceed perturbatively in $\tau $, evaluating the
second moment on the right hand side, using the approximation at previous
order. Again, this will only account for a modification $A\mapsto Ad_{k}$,
with 
\begin{equation}
d_{k}=\sum_{j=0}^{k}\left( j+1\right) \tau ^{j}\left[ \int_{-\alpha
}^{\alpha }du\rho (u)u^{2}-\frac{1}{12}\right] ^{j},  \label{eq:recursiveDk}
\end{equation}%
where the second moment of the distribution is evaluated at order $k-1$.

We directly treat the problem at a generic order $k$, knowing that $d_{0}=1$%
, and hence the initial step of our procedure corresponds to the result of 
\cite{Douglas:1993iia}. We define a complex function\footnote{%
Cfr., for instance, \cite[Ch.10]{Pipkin:1991} or \cite[Ch.11]{Muskhe:1977}
for a review of the procedure.}%
\begin{equation}
\Phi (z)=\int \frac{\varphi (u)}{u-z}du,  \label{eq:intdefPhi}
\end{equation}%
for $z\notin \left[ -\alpha ,\alpha \right] $. On one hand, when $z$
approaches a real number $h\in \left[ -\alpha ,\alpha \right] $, we can
write 
\begin{equation}
\Phi _{+}\left( h\right) -\Phi _{-}\left( h\right) =2\pi \mathrm{i}\varphi
(h)1\!\!1_{U}(h),  \label{eq:Phiequalsphi}
\end{equation}%
where $\Phi _{\pm }(h):=\lim_{\varepsilon \rightarrow 0}\Phi \left( h\pm 
\mathrm{i}\varepsilon \right) $ and $1\!\!1_{U}$ is the characteristic
function of the set 
\begin{equation*}
U:=\left[ -\alpha ,-\beta \right] \cup \left[ \beta ,\alpha \right] \equiv %
\left[ -\alpha ,\alpha \right] \setminus \left( -\beta ,\beta \right) .
\end{equation*}%
Once we obtain a complex solution to the saddle point equation %
\eqref{eq:SPE2}, we can recover the $\varphi $ by evaluating the left hand
side of \eqref{eq:Phiequalsphi} as the discontinuity at the branch cut of
the complex solution for $h\in \left[ -\alpha ,\alpha \right] $. Such a
complex solution is: 
\begin{equation}
\Phi (z)=-\frac{1}{2\pi \mathrm{i}}\sqrt{\left( \alpha ^{2}-z^{2}\right)
\left( \beta ^{2}-z^{2}\right) }\oint_{\gamma _{U}}du\frac{\frac{Ad_{k}}{2}%
u+\log \left( \frac{u-\beta }{u+\beta }\right) }{\left( u-z\right) \sqrt{%
\left( \alpha ^{2}-u^{2}\right) \left( \beta ^{2}-u^{2}\right) }},
\end{equation}%
for some path $\gamma _{U}$ in the complex plane around the cut $U$. After
an adequate deformation of the contour integral, one gets: 
\begin{equation}
\Phi (z)=-\frac{Ad_{k}}{2}z-\log \left( \frac{z-\beta }{z+\beta }\right) -%
\sqrt{\left( \alpha ^{2}-z^{2}\right) \left( \beta ^{2}-z^{2}\right) }%
\int_{-\beta }^{\beta }\frac{du}{\left( u-z\right) \sqrt{\left( \alpha
^{2}-u^{2}\right) \left( \beta ^{2}-u^{2}\right) }}.  \label{eq:Phifullexpr}
\end{equation}%
The first two terms are obtained from the residue theorem, and the last one
accounts for the branch cut of the logarithm along $\left[ -\beta ,\beta %
\right] $. When $z$ approaches the real axis, the logarithm has a
discontinuity of $2\pi \mathrm{i}$ if $z\rightarrow h\in \left[ -\beta
,\beta \right] $, and has no discontinuity out of that interval, while the
third term is discontinuous in all $z\in \left[ -\alpha ,\alpha \right] $.
Thus we get: 
\begin{equation}
\Phi _{+}\left( h\right) -\Phi _{-}\left( h\right) =-2\pi \mathrm{i}1\!\!1_{%
\left[ -\beta ,\beta \right] }-2\mathrm{i}\mathrm{sign}(h)\sqrt{\left(
\alpha ^{2}-h^{2}\right) \left( h^{2}-\beta ^{2}\right) }\int_{-\beta
}^{\beta }\frac{du}{\left( u-h\right) \sqrt{\left( \alpha ^{2}-u^{2}\right)
\left( \beta ^{2}-u^{2}\right) }}.
\end{equation}%
The sign function appears because, for $h\in \left[ \beta ,\alpha \right] $,
one approaches the branch cut of the square root from the proper direction,
i.e. $\Phi _{+}-\Phi _{-}$ corresponds to \textquotedblleft
counter-clockwise minus clockwise\textquotedblright . For $h\in \left[
-\alpha ,-\beta \right] $, instead, the branch cut is approached from the
converse direction.

Therefore, comparing with \eqref{eq:Phiequalsphi}, one arrives to: 
\begin{equation}  \label{eq:twocutrho}
\begin{aligned} \rho (h) & = \varphi (h) 1\!\!1_{U} + 1\!\!1_{\left[-\beta,
\beta\right]} \\ & = \frac{1}{\pi} \mathrm{sign} (h ) \sqrt{ \left( \alpha^2
- h^2 \right) \left( h^2 - \beta^2 \right) } \int_{- \beta} ^{\beta} \frac{
d u}{\left( h - u \right) \sqrt{\left( \alpha^2 - u^2 \right) \left( \beta^2
- u^2 \right)}} . \end{aligned}
\end{equation}
It is easy to check that eigenvalue distribution in a positive function of $%
h $ in all $\left[ - \alpha, \alpha \right]$ and is identically $1$ in the
interval $\left[ - \beta, \beta \right]$.

Until this point we simply reproduced the procedure of \cite{Douglas:1993iia}%
, which applies also for our generalized case. Notice that the eigenvalue
distribution $\rho $ apparently does not yield an explicit dependence on $%
Ad_{k}$; nevertheless, the parameters $\alpha ,\beta $ will depend on it,
and so will do $\rho $. The boundaries $\alpha ,\beta $ can be fixed by the
asymptotic expansion of $\Phi (z)$, for instance by comparison between %
\eqref{eq:Phifullexpr} and the definition \eqref{eq:intdefPhi}. From this
latter we have: 
\begin{equation}
\begin{aligned} \Phi (z) & = - \frac{1}{z} \left( \int_U du \varphi (u) +
\frac{1}{z^2} \int_U du \varphi (u) u^2 + \dots \right) \\ & = - \frac{1}{z}
\left[ \left(1 - 2 \beta \right) + \frac{1}{z^2} \left( \int_{- \alpha}
^{\alpha} du \rho (u) u^2 - \frac{2}{3} \beta^3 \right) + \dots \right] .
\end{aligned}
\end{equation}%
On the other hand, the explicit expression \eqref{eq:Phifullexpr} implies: 
\begin{equation}
\begin{aligned} \Phi (z) & = - \frac{A d_k }{2} z - \log \left( 1 - \frac{2
\beta}{z} + \dots \right) + z \left( 1 - \frac{ \alpha^2 + \beta^2 }{2 z^2 }
+ \dots \right) \int_{- \beta } ^{\beta} du \frac{ \left[ 1 +
\frac{u^2}{z^2} + \frac{ u^4 }{z^4} + \dots \right] }{ \sqrt{\left( \alpha^2
- u^2 \right) \left( \beta^2 - u^2 \right)}} \\ & = z \left[ - \frac{A d_k
}{2} z + \int_{- \beta } ^{\beta} \frac{ du }{ \sqrt{\left( \alpha^2 - u^2
\right) \left( \beta^2 - u^2 \right)}} \right] + \frac{1}{z} \left[2 \beta +
\int_{- \beta} ^{\beta} du \frac{ u^2 - \frac{\alpha^2 + \beta^2}{2} }{
\sqrt{\left( \alpha^2 - u^2 \right) \left( \beta^2 - u^2 \right)}} \right] +
\mathcal{O} \left( \frac{1}{z^3} \right) \end{aligned}
\end{equation}%
The comparison at $\mathcal{O}(z)$ imposes the constraint 
\begin{equation}
\int_{-\beta }^{\beta }\frac{du}{\sqrt{\left( \alpha ^{2}-u^{2}\right)
\left( \beta ^{2}-u^{2}\right) }}=\frac{Ad_{k}}{2}\quad \Longrightarrow
\quad \alpha =\frac{4}{Ad_{k}}K\left( \frac{\beta }{\alpha }\right) ,
\label{eq:constraint1Phi}
\end{equation}%
where $K\left( \cdot \right) $ is the complete elliptic integral of first
kind. Analogously, from comparison at $\mathcal{O}(z^{-1})$ one gets the
constraint: 
\begin{equation}
\begin{aligned} & 2 \beta - \frac{\alpha^2 + \beta^2}{2} \int_{- \beta }
^{\beta} \frac{ du }{ \sqrt{\left( \alpha^2 - u^2 \right) \left( \beta^2 -
u^2 \right)}} + \int_{- \beta } ^{\beta} \frac{ u^2 du }{ \sqrt{\left(
\alpha^2 - u^2 \right) \left( \beta^2 - u^2 \right)}} = - 1 + 2 \beta \\ &
\Longrightarrow \quad K \left( \frac{\beta}{\alpha} \right) \left[ 2 E
\left( \frac{\beta}{\alpha} \right) - \left( 1 - \frac{\beta^2}{\alpha^2}
\right) K \left( \frac{\beta}{\alpha} \right) \right] = \frac{A d_k}{4} ,
\end{aligned}  \label{eq:constraint2Phi}
\end{equation}%
where $E\left( \cdot \right) $ is the complete elliptic integral of second
kind, and we plugged in \eqref{eq:constraint1Phi} to simplify the
expression. The result is clearly the same as \cite{Douglas:1993iia}, up to
a rescaling $A\mapsto Ad_{k}$.

As we are interested in knowing the second moment of the distribution $\rho $%
, we may use $\mathcal{O}(z^{-3})$ of the expansion above to obtain the
dependence of the integral expression on the other parameters. It leads to: 
\begin{equation}
\begin{aligned} \int_{- \alpha} ^{\alpha} du \rho (u) u^2 & = \int_{- \beta}
^{ \beta} d u \frac{ - u^4 + \frac{ \alpha^2 + \beta^2 }{2} u^2 + \frac{
\left( \alpha^2 - \beta^2 \right)^2 }{8} }{ \sqrt{ \left( \alpha^2 - u^2
\right) \left( \beta^2 - u^2 \right) } } \\ & = \frac{ \left( \alpha^2 -
\beta^2 \right)^2 }{16} A d_k + \alpha \left( \alpha^2 + \beta^2 \right)
\left[ K \left( \frac{ \beta }{ \alpha } \right) - E \left( \frac{ \beta }{
\alpha } \right) \right] \\ & - \frac{ 2 }{3} \alpha^3 \left[ \left( 2 +
\frac{\beta^2 }{\alpha^2} \right) K \left( \frac{ \beta}{\alpha}\right) - 2
\left( 1 + \frac{\beta^2 }{\alpha^2} \right) E \left( \frac{
\beta}{\alpha}\right) \right] . \end{aligned}  \label{eq:constraint3Phi}
\end{equation}

We can use the properties of the elliptic integrals to extract information
about the dependence on $Ad_{k}$. In particular, from the first two
conditions \eqref{eq:constraint1Phi}-\eqref{eq:constraint2Phi} we get that,
for $Ad_{k}\rightarrow \pi ^{2}$, one recovers the same parameters as
approaching the critical point from below, that is, $\left( \alpha =2/\pi
,\beta =0\right) $. More specifically, for $Ad_{k}$ close to $\pi ^{2}$, we
may approximate the elliptic integrals, and obtain the first terms of the
expansion of $\alpha $ and $\beta $ around $Ad_{k}=\pi ^{2}$:%
\begin{equation}
\begin{aligned} \alpha & = \frac{1}{\pi} \left[ 2 - \frac{ A d_{k} - \pi^2
}{\pi^2} + \frac{ 5}{4} \left( \frac{ A d_{k} - \pi^2 }{\pi^2} \right)^2 +
\dots \right], \\ \beta & = \frac{1}{\pi} \left[ 2 \sqrt{2} \left( \frac{
\left( A d_{k} - \pi^2 \right) }{\pi^2} \right)^{\frac{1}{2}} - \frac{15}{4
\sqrt{2} } \left( \frac{ A d_{k} - \pi^2 }{\pi^2} \right)^{\frac{3}{2}} +
\dots \right]. \end{aligned}
\end{equation}%
Moreover, concerning the second moment, approximated close to the critical
point, we get: 
\begin{equation}
\int_{-\alpha }^{\alpha }du\rho (u)u^{2}=\frac{1}{\pi ^{2}}\left[ 1-\frac{%
Ad_{k}-\pi ^{2}}{\pi ^{2}}+3\left( \frac{Ad_{k}-\pi ^{2}}{\pi ^{2}}\right)
^{2}+\dots \right] .  \label{eq:approx2ndmomentum}
\end{equation}

At this point, we are able to determine the full nonperturbative expression
for the eigenvalue density $\rho (h)$, close enough to the critical point.
That is: on one hand, we have a formal recursive expression for the
coefficients $d_{k}$ of the perturbative expansion in $\tau $ at strong
coupling, while on the other hand, if we want to determine the order of the
phase transition, we need to know an explicit expression for the dependence
of the eigenvalue density on the parameters $A$ and $\tau $. Notice however,
that local information close to the critical point is enough to
characterize the phase transition. For these reasons, we look for a full
(nonperturbative) solution, approximating close to the critical point. The
solution we will find will be only valid up to order $(Ad_{\infty }-\pi
^{2})^2$.

The formal limit $k\rightarrow \infty $ of this expression %
\eqref{eq:recursiveDk} leads to the equation: 
\begin{equation}
d_{\infty }=\left[ 1-\tau \left( \int_{- \alpha} ^{\alpha} du \rho (u) u^2 -%
\frac{1}{12}\right) \right] ^{-2},  \label{eq:defDinf}
\end{equation}%
and the approximated solution close to the critical point is found plugging
expression \eqref{eq:approx2ndmomentum}, obtaining: 
\begin{equation}
d_{\infty } \approx \left[ 1-\tau \left[ \frac{1}{\pi^2}\left( 1 - \frac{ A
d_{\infty} - \pi^2 }{\pi^2} \right) -\frac{1}{12} \right] \right] ^{-2} ,
\end{equation}
which again admits only one solution compatible with $\lim_{\tau \rightarrow
0}d_{\infty }\left( A,\tau \right) =1$. As a side remark, we highlight that
the defining equation for $d_{\infty}$ starts to differ from the one for $%
b_{\infty}$ only at order $(A d_{\infty} - \pi^2)^2$, implying that $%
b_{\infty}$ and $d_{\infty}$ will coincide up to the first derivative when
evaluated at the critical point $A_{cr}$ (same $1$-jet at $A_{cr}$).

\subsection{Third order phase transition}

In this subsection, we study the free energy from the point of view of small
and large area, that is $A<A_{cr}\left( \tau \right) $ and $A>A_{cr}\left(
\tau \right) $ respectively, with the aim to determine the order of the
phase transition. The free energy of the system is defined as: 
\begin{equation}
\mathcal{F}_{N}\left( A,\tau \right) =-\frac{1}{N^{2}}\mathcal{Z}_{N}\left(
A,\tau \right) .
\end{equation}%
In the large $N$ limit the derivative with respect to the control parameter $%
A$ is given by: 
\begin{equation}
\frac{\partial \mathcal{F}}{\partial A}=\frac{1}{2}\sum_{j=0}^{\infty }\tau
^{j}\left[ \int du\rho (u)u^{2}-\frac{1}{12}\right] ^{j+1}.
\end{equation}%
Before passing to the direct evaluation, we notice that: 
\begin{equation}
\frac{\partial \mathcal{F}}{\partial A}=F_{\mathrm{DK}}^{\prime }\left(
Ac_{\infty }\right) \sum_{j=0}^{\infty }\tau ^{j}\left[ \int du\rho (u)u^{2}-%
\frac{1}{12}\right] ^{j},
\end{equation}%
where by ${F}_{\mathrm{DK}}^{\prime }(A)$ we mean the first derivative of
the free energy obtained by Douglas and Kazakov \cite{Douglas:1993iia}, and $%
c_{\infty }$ is a shorthand: 
\begin{equation*}
c_{\infty }=%
\begin{cases}
b_{\infty },\quad A<A_{cr}(\tau ); \\ 
d_{\infty },\quad A>A_{cr}(\tau ).%
\end{cases}%
\end{equation*}%
Therefore 
\begin{equation}
\frac{\partial \mathcal{F}}{\partial A}={F}_{\mathrm{DK}}^{\prime
}(Ac_{\infty })\left[ 1-\tau \left( \int du\rho (u)u^{2}-\frac{1}{12}\right) %
\right] ^{-1}.
\end{equation}%
Taking advantage of the defining equation \eqref{eq:defBinf} and %
\eqref{eq:defDinf} for $b_{\infty }$ and $d_{\infty }$ respectively, we can
rewrite: 
\begin{equation}
\frac{\partial \mathcal{F}}{\partial A}={F}_{\mathrm{DK}}^{\prime
}(Ac_{\infty })\sqrt{c_{\infty }}.
\end{equation}%
When $A<A_{cr}$, the latter expression is calculated using the distribution
at small coupling:%
\begin{equation}
\frac{\partial \mathcal{F}}{\partial A}\bigg{\vert}_{A<A_{cr}}=\frac{\sqrt{%
b_{\infty }}}{2}\left( \frac{1}{Ab_{\infty }}-\frac{1}{12}\right) =\frac{%
\sqrt{b_{\infty }}}{2}\left[ \frac{1}{\pi ^{2}}\left( 1-\frac{Ab_{\infty
}-\pi ^{2}}{\pi ^{2}}+\left( \frac{Ab_{\infty }-\pi ^{2}}{\pi ^{2}}\right)
^{2}+\dots \right) -\frac{1}{12}\right] .
\end{equation}%
Analogously, in the strong coupling phase $A>A_{cr}(\tau )$ we ought to use
the eigenvalue distribution at strong coupling, which, approximating close
to the critical point, provides the expression: 
\begin{equation*}
\frac{\partial \mathcal{F}}{\partial A}\bigg{\vert}_{A>A_{cr}}=\frac{\sqrt{%
d_{\infty }}}{2}\left[ \frac{1}{\pi ^{2}}\left( 1-\frac{Ad_{\infty }-\pi ^{2}%
}{\pi ^{2}}+3\left( \frac{Ad_{\infty }-\pi ^{2}}{\pi ^{2}}\right) ^{2}+\dots
\right) -\frac{1}{12}\right] .
\end{equation*}%
By construction of the two-cut solution, we know that: 
\begin{equation}
Ab_{\infty }\xrightarrow{ \ A \to A_{cr} ^{-} \ }\pi ^{2}\xleftarrow{ \ A
\to A_{cr} ^{+} \ }Ad_{\infty }
\end{equation}%
which guarantees $\frac{\partial \mathcal{F}}{\partial A}$ is continuous at
the critical point, thus the transition is at least of second order. We in
fact have that: 
\begin{equation}
\begin{aligned} \frac{\partial \mathcal{F}}{\partial
A}\bigg{\vert}_{A>A_{cr}}-\frac{\partial \mathcal{F}}{\partial
A}\bigg{\vert}_{A<A_{cr}} & = \frac{1}{2} \left( \frac{1}{\pi^2} -
\frac{1}{12} \right)\left( \sqrt{d_{\infty}} - \sqrt{b_{\infty}} \right) \\
& - \frac{1}{2 \pi^2} \left[ \sqrt{d_{\infty}} \left( \frac{ A d_{\infty} -
\pi^2 }{\pi^2} \right) - \sqrt{b_{\infty}} \left( \frac{ A b_{\infty} -
\pi^2 }{\pi^2} \right) \right] \\ & + \frac{1}{2 \pi^2} \left[ 3
\sqrt{d_{\infty}} \left( \frac{ A d_{\infty} - \pi^2 }{\pi^2} \right)^2 -
\sqrt{b_{\infty}} \left( \frac{ A b_{\infty} - \pi^2 }{\pi^2} \right)^2
\right] + \dots . \end{aligned}
\end{equation}%
Taking a further derivative with respect to $A$ we get: 
\begin{equation}
\begin{aligned} \frac{ \partial^2 \fre }{\partial A^2} \bigg{\vert}_{A >
A_{cr} } - \frac{ \partial^2 \fre }{\partial A^2} \bigg{\vert}_{A < A_{cr} }
& = \frac{1}{4} \left( \frac{1}{\pi^2} - \frac{1}{12} \right)\left[ \frac{
d_{\infty} ^{\prime} }{\sqrt{d_{\infty}}} - \frac{ b_{\infty} ^{\prime}
}{\sqrt{b_{\infty}}} \right] \\ & - \frac{1}{4 \pi^2} \left[ \frac{
d_{\infty} ^{\prime} }{\sqrt{d_{\infty}}} \left( \frac{ A d_{\infty} - \pi^2
}{\pi^2} \right) - \frac{ b_{\infty} ^{\prime} }{\sqrt{b_{\infty}}} \left(
\frac{ A b_{\infty} - \pi^2 }{\pi^2} \right) \right] \\ & - \frac{1}{2
\pi^4} \left[ \sqrt{d_{\infty}} \left( d_{\infty} + A d_{\infty} ^{\prime}
\right) - \sqrt{b_{\infty}} \left( b_{\infty} + A b_{\infty} ^{\prime}
\right) \right] \\ & + \frac{1}{4 \pi^2} \left[ 3 \frac{ d_{\infty}
^{\prime} }{\sqrt{d_{\infty}}} \left( \frac{ A d_{\infty} - \pi^2 }{\pi^2}
\right)^2 - \frac{ b_{\infty} ^{\prime} }{\sqrt{b_{\infty}}} \left( \frac{ A
b_{\infty} - \pi^2 }{\pi^2} \right)^2 \right] \\ & + \frac{1}{\pi^4} \left[
\sqrt{d_{\infty}} \left( d_{\infty} + A d_{\infty} ^{\prime} \right) \left(
\frac{ A d_{\infty} - \pi^2 }{\pi^2} \right) - \sqrt{b_{\infty}} \left(
b_{\infty} + A b_{\infty} ^{\prime} \right)\left( \frac{ A b_{\infty} -
\pi^2 }{\pi^2} \right) \right] + \dots \end{aligned}
\end{equation}%
Is is straightforward to see that the second, fourth and fifth term vanish
at the critical point. The first and third term are more subtle, because the
involve derivatives of the coefficients $b_{\infty },d_{\infty }$. However,
we have that both coefficients are defined by formally the same expression,
but using the eigenvalue distribution at small or large coupling
respectively. Regarding $b_{\infty }$, we can evaluate its derivative using %
\eqref{eq:defBinf}:%
\begin{equation*}
b_{\infty }^{\prime }=-\frac{2\tau }{\pi ^{2}}\frac{b_{\infty }^{5/2}}{%
1-Ab_{\infty }^{3/2}}+\dots ,
\end{equation*}%
where the dots represent term that vanish at the critical point. The same
can be done for $d_{\infty }$ using \eqref{eq:defDinf}, to obtain:%
\begin{equation*}
d_{\infty }^{\prime }=-\frac{2\tau }{\pi ^{2}}\frac{d_{\infty }^{5/2}}{%
1-Ad_{\infty }^{3/2}}+\dots .
\end{equation*}%
Hence we infer that the first derivatives $b_{\infty }^{\prime }$ and $%
d_{\infty }^{\prime }$ of the coefficients coincide at the critical point
(this fails to be true for higher derivatives). Consequently, the phase
transition is again of third order.

It is remarkable that the $T\overline{T}$-deformation introduced a
nontrivial dependence on the coupling $A$, but in such a way that it does
not affect the order of the phase transition.

\section{Instanton analysis}

\label{sec:instantons}

Gross and Matytsin presented evidence for the phase transition to be
triggered by instantons \cite{Gross:1994mr}. The existence of a phase
transition can be closely related to the discreteness of the matrix model %
\eqref{HK} \cite{Gross:1994mr,Jafferis:2005jd,Szabo:2013vva}. In \cite%
{Minahan:1993tp}, Yang--Mills theory on the sphere is described in terms of $%
N$ nonrelativistic free fermions on $\cs^1$, with instantons corresponding
to different winding numbers for fermions at a given position, and the phase
transition occurs due to the condensation of fermions in (discrete) momentum
space, so again the discreteness turned out to be essential to permit a
phase transition. From this general argument, and taking into account
expression \eqref{eq:ttZYM2}, the statement is expected to hold also in the $%
T\overline{T}$-deformed version of two-dimensional Yang--Mills, as the
effect of the deformation, at the level of the matrix model, is to replace
the discrete Gaussian weight with a discrete weight whose potential has also
additional multitrace contributions. Thus, in this section we look at the
role played by instantons in the phase transition.

\subsection{Instantons in the undeformed theory}

By instanton, we mean a solution of the classical Yang--Mills equation of
motion which is gauge inequivalent to the trivial one. Those solutions are
in one to one correspondence with collections of $N$ monopole charges:%
\begin{equation*}
\ell :=\left( \ell _{1},\dots ,\ell _{N}\right) \in \Zint^{N}.
\end{equation*}%
The action for a given classical configuration is: 
\begin{equation}
S_{inst}\left( \ell \right) =\frac{N}{2A}\sum_{j=1}^{N}\left( 2\pi \ell
_{j}\right) ^{2},
\end{equation}%
and the partition function splits into the sum of contributions from
instanton sectors: 
\begin{equation}
\mathcal{Z}_{N}^{(\mathrm{YM})}=\sum_{\ell \in \Zint^{N}}w\left( \ell
\right) e^{-S_{inst}\left( \ell \right) }.
\end{equation}%
This is the content of Witten's result \cite{Witten:1991we,Witten:1992xu}
extending the Duistermaat--Heckman theorem. From the point of view of the
Abelianization procedure \cite{Blau:1993tv}, each $\ell _{j}$ is the first
Chern class of a $U(1)$-bundle. See \cite%
{Blau:1993tv,Blau:1993hj,Cordes:1994fc} for more details. Later on, this
viewpoint was the one adopted in \cite{Jafferis:2005jd} to estimate the
instanton contributions in the case of $q$-deformed Yang--Mills theory on $%
\cs^2$.

A practical difficulty is to evaluate the weights $w \left( \ell \right)$,
which was done in \cite{Minahan:1993tp} (and in \cite{Caporaso:2005ta} for
the $q$-deformed case), through the method of Poisson resummation. In \cite%
{Gross:1994mr}, the contribution of the single-monopole sector $\ell =
\left( 1, 0, \dots, 0 \right)$ was calculated, showing that this correction
to the saddle point approximation at large $N$ is exponentially suppressed, 
\begin{equation*}
\frac{ w \left( 1 , 0 , \dots, 0 \right) e^{- S_{inst} \left( 1 , 0 , \dots,
0 \right) } }{ w \left( 0 , \dots, 0 \right) e^{- S_{inst} \left( 0 , \dots,
0 \right) } } \propto e^{- \frac{2 \pi^2}{A} N \gamma \left( \frac{A}{\pi^2}
\right) } ,
\end{equation*}
where the function $\gamma \left( \cdot \right)$ is the one introduced by
Gross and Matytsin \cite{Gross:1994mr} and is given by 
\begin{equation}  \label{eq:GMgamma}
\gamma (x) = \sqrt{ 1 - x} - \frac{x}{2} \log \left( \frac{ 1 + \sqrt{1 - x }%
}{1 - \sqrt{1 - x}} \right) .
\end{equation}
In particular, as $\gamma \left( x\right) >0$ for $0<x<1$ and $\gamma \left(
1\right) =0$, in the small coupling phase contributions from instanton
sectors are exponentially suppressed at large $N$, but they become more and
more relevant as the critical point is approached. In this sense, the weight 
$w \left( \ell \right)$ acts as a counterpart of the Boltzmann factor $e^{-
S \left( \ell \right) }$, and at the critical point those two contributions
are exactly balanced.

\subsection{Instantons in the $T \overline{T}$-deformed theory}

The evaluation of the full instanton expansion for the deformed theory would
correspond to find an explicit expression of the form: 
\begin{equation*}
\mathcal{Z}_{N} \left( A,\tau \right) =\sum_{\ell \in \Zint^{N}}w\left( \ell
\right) e^{-S_{inst}\left( \ell \right) },
\end{equation*}%
where now the weights and the action include the effects of the deformation
by the $T\overline{T}$-operator. We have found strong evidence that this can
be done in an analytic way, but the weights obtained are not very
enlightening and unsuitable for the purpose of this section. Instead, we
will only look at the first instanton correction, and how it affects the
model. A discussion on the Poisson resummation for the full explicit
expression is relegated to Appendix \ref{sec:InstantonPoissonResum}.

As a first step, the partition function can be rewritten as a sum over
Fourier transforms of contributions from each representation: 
\begin{equation}
\mathcal{Z}_{N} \left( A,\tau \right) =\sum_{\ell \in \Zint^{N}} Z_{\ell },
\end{equation}%
where each instanton sector contributes as: 
\begin{equation}
Z_{\ell }=\int_{ \R ^{N}}\prod_{i=1}^{N}dh_{i}e^{-\mathrm{i}2\pi
\sum_{i=1}^{N}\ell _{i}h_{i}}\prod_{i<j}\left( \frac{h_{i}-h_{j}}{j-i}%
\right) ^{2}e^{-\frac{A}{2N}\sum_{j=0}^{\infty }\left( \frac{\tau }{N^{3}}%
\right) ^{j}\left( \sum_{i=1}^{N}h_{i}^{2}-\frac{N\left( N^{2}+2\right) }{12}%
\right) ^{j+1}}.
\end{equation}%
Consider the single-monopole sector corresponding to $\ell =\left( \ell
_{1},0,\dots ,0\right) $ (we will eventually set $\ell _{1}=\pm 1$). It
contributes to the partition function as: 
\begin{equation*}
Z_{\left( \ell _{1},0,\dots ,0\right) }=\int_{ \R^{N}}%
\prod_{i=1}^{N}dh_{i}e^{-N^{2}S_{\ell _{1}}[h]},
\end{equation*}%
where 
\begin{equation}
S_{\ell _{1}}[h]=-\frac{2}{N^{2}}\sum_{i<j}\log \left( \frac{h_{i}-h_{j}}{j-i%
}\right) +\frac{A}{2N^{3}}\sum_{j=0}^{\infty }\left( \frac{\tau }{N^{3}}%
\right) ^{j}\left( \sum_{i=1}^{N}h_{i}^{2}-\frac{N\left( N^{2}+2\right) }{12}%
\right) ^{j+1}+\frac{2\pi \mathrm{i}}{N^{2}}\ell _{1}h_{1}.
\end{equation}%
This means that the correction to the large $N$ action, with respect to the
vacuum sector, is of $\mathcal{O}(N^{-1})$. This implies that, at large $N$,
we can perform $N-1$ integrals using the saddle point approximation for the
eigenvalue distribution, and eventually treat the integration over $h_{1}$
separately \cite{Gross:1994mr,Jafferis:2005jd}. Notice that the saddle point
for $h_{1}$ will be, in general, complex, due to the purely imaginary
``Fourier interaction'' with $\ell _{1}$. Nevertheless, as we are interested
in the suppressing factor, we will avoid the technicalities involved in
determining the imaginary part of the instanton contribution.

Taking the large $N$ limit, we have: 
\begin{equation}
Z_{\left( \ell _{1},0,\dots ,0\right) }=\mathcal{C}^{N-1}\int_{-\infty
}^{\infty }dh_{1}e^{-NS_{\ell _{1}}[h_{1}]},
\end{equation}%
where $\mathcal{C}$ is the integral over any variable $h_{2},\dots ,h_{N}$,
calculated with the saddle point approximation, and the effective action for
the scaled variable $h=\frac{h_{1}}{N}$ is: 
\begin{equation}
S_{\ell _{1}}[h]=-2\int du\rho (u)\log \left( h-u\right) -2+\frac{A}{2}%
\sum_{j=0}^{\infty }\tau ^{j}\left( h^{2}-\frac{1}{12}\right) ^{j+1}+2\pi 
\mathrm{i}\ell _{1}h,
\end{equation}%
and the density $\rho $ is the one obtained in \eqref{eq:FullWignerSmallA}
for small coupling $A<A_{cr}$. The saddle point for the effective action is
given by:%
\begin{equation*}
2\mathrm{P}\int du\frac{\rho (u)}{h-u}+Ah\sum_{j=0}^{\infty }\left(
j+1\right) \tau ^{j}\left( h^{2}-\frac{1}{12}\right) ^{j}+2\pi \mathrm{i}%
\ell _{1}=0
\end{equation*}%
which, using the fact that $\rho $ satisfies the saddle point for $\ell
_{1}=0$, simplifies into: 
\begin{equation*}
2\pi \mathrm{i}\left( \rho (h)+\ell _{1}\right) =0.
\end{equation*}%
This leads to the saddle point: 
\begin{equation*}
h^{2}=\left( \frac{2}{Ab_{\infty }}\right) ^{2}\left( Ab_{\infty }-\pi
^{2}\ell _{1}\right) ,
\end{equation*}%
which, since the analysis is being brought on in the small coupling phase $%
Ab_{\infty }<\pi ^{2}$, gives a purely imaginary saddle point: 
\begin{equation}
h=\frac{2\pi \mathrm{i}}{Ab_{\infty }}\mathrm{sign}(\ell _{1})\sqrt{\ell
_{1}^{2}-\frac{Ab_{\infty }}{\pi ^{2}}}.
\end{equation}%
The result is exactly the same obtained in the undeformed case \cite%
{Gross:1994mr}, up to a rescaling $A\mapsto Ab_{\infty }$. In particular: 
\begin{equation}
\frac{Z_{\left( 1,0,\dots ,0\right) }}{Z_{\left( 0,\dots ,0\right) }}=%
\mathcal{C}^{\prime }e^{-N\frac{2\pi ^{2}}{Ab_{\infty }}\gamma \left( \frac{%
Ab_{\infty }}{\pi ^{2}}\right) },
\end{equation}%
where $\mathcal{C}^{\prime }$ is an overall constant and $\gamma \left(
\cdot \right) $ is the function \eqref{eq:GMgamma}. Hence, the same
conclusions of the undeformed case \cite{Gross:1994mr} hold: from the
perspective of the small coupling expansion, the phase transition is
triggered by instantons.

A more thorough analysis of the single-monopole instanton correction in the $%
T\overline{T}$-deformed case, shows that the relevance of instantons
increases with $\tau $, for $\tau < \frac{12 \pi^2}{12-\pi^2}$, and there is no suppression at all for $\tau \ge \frac{12 \pi^2}{12-\pi^2}$. This explains why the phase transition occurs
earlier in the deformed case, that is $A_{cr}(\tau )\leq A_{cr}(0)$: indeed,
the function $\frac{1}{Ab_{\infty }}\gamma \left( \frac{Ab_{\infty }}{\pi
^{2}}\right) $ decreases faster than the function $\frac{1}{A}\gamma \left( 
\frac{A}{\pi ^{2}}\right) $, hence the first instanton sector becomes
relevant at a lower value of $A$, in comparison to the undeformed case. This
is presented in Figures \ref{fig:gammaatt01} and \ref{fig:gammaatt05}.

{\begin{figure}[htb]
\centering
\includegraphics[width=0.43\textwidth]{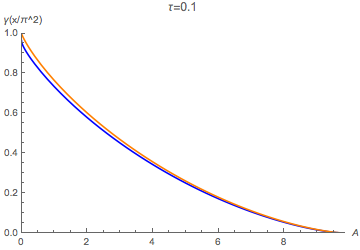}\hspace{0.1\textwidth} %
\includegraphics[width=0.43\textwidth]{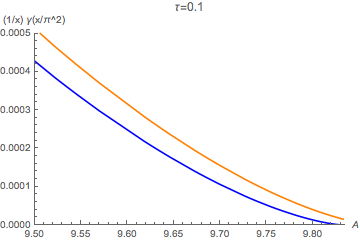}
\caption{On the left: a comparison of the function $\protect\gamma \left( 
\frac{x}{\protect\pi^2} \right)$ for the deformed (blue) and undeformed
(orange) case. On the right: a zoom on the tail of the function $\frac{1}{x} 
\protect\gamma \left( \frac{x}{\protect\pi^2} \right)$ for the deformed
(blue) and undeformed (orange) case. The plots are at $\protect\tau = 0.1$.}
\label{fig:gammaatt01}
\end{figure}

\begin{figure}[htb]
\centering
\includegraphics[width=0.43\textwidth]{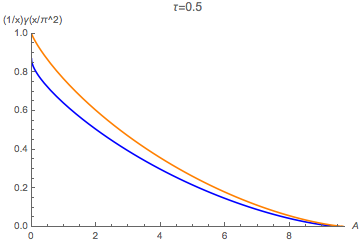}\hspace{0.1\textwidth} %
\includegraphics[width=0.43\textwidth]{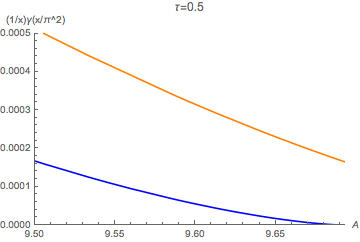}
\caption{On the left: a comparison of the function $\protect\gamma \left( 
\frac{x}{\protect\pi^2} \right)$ for the deformed (blue) and undeformed
(orange) case. On the right: a zoom on the tail of the function $\frac{1}{x} 
\protect\gamma \left( \frac{x}{\protect\pi^2} \right)$ for the deformed
(blue) and undeformed (orange) case. The plots are at $\protect\tau = 0.5$.}
\label{fig:gammaatt05}
\end{figure}
}

\section{Outlook}

It is worth mentioning a few possible open problems, from the point of view
of classical results involving two-dimensional Yang--Mills theory on the
sphere. One is the study of Wilson loops, a well-known solvable problem in
the undeformed theory. Some results in this regard are immediately available
by using some of the results above. Recall that already the zero instanton
sector of ordinary two-dimensional Yang--Mills theory is well-known to
capture the vacuum expectation value of Wilson loops in $4d$ $\mathcal{N}=4$
supersymmetric Yang--Mills theory \cite{Giombi:2009ds,Giombi:2009ek}. The
zero instanton sector of the ordinary theory is described by a Gaussian
matrix model and hence, the Wilson loop average in that ensemble is the
well-known \cite{Bassetto:1998sr,Drukker:2000rr} 
\begin{equation}
\left\langle W_{0}\right\rangle =\frac{1}{N}\exp \left( -g^{2}\frac{%
A_{1}A_{2}}{2A}\right) L_{N-1}^{1}\left( g^{2}\frac{A_{1}A_{2}}{A}\right) ,
\label{W0}
\end{equation}%
where $L_{N-1}^{1}(x)$ is a Laguerre polynomial.

We have obtained that the density of states in the weak coupling phase is
given by a Wigner semicircle law with a rescaled parameter $A\mapsto
Ab_{\infty }$. Hence the large $N$ Wilson loop in the zero-instanton sector
of the $T\overline{T}$-deformed theory will be given by a rescaled version
of the large $N$ limit of (\ref{W0}), which is given by a Bessel function,
and has a celebrated dual holographic description \cite{Drukker:2000rr}.
Thus, it is interesting to carry out the analysis of the Wilson loops in the 
$T\overline{T}$-deformed theory and look for potential implications in
supersymmetric gauge theory.

It is also worth mentioning that some results on Wilson loops in $2d$
Yang--Mills theory on the plane, in part to further understand the
Duurhus--Olensen transition \cite{Durhuus:1980nb}, precisely make use of the
existence of a Burgers equation for certain generating functions (in the
form of characteristic polynomials) of the Wilson loops \cite%
{Blaizot:2008nc,Neuberger:2008ti}. Since this property follows from the heat
kernel propagator, which is preserved under the deformation, such methods
and studies will seemingly translate to the deformed case as well.

Two-dimensional Yang--Mills theory has many other interesting aspects that
could now be re-analyzed under the prism of the $T\overline{T}$-deformation
of the theory. Clear examples are the factorization of the theory in chiral
and anti-chiral sectors and the corresponding Gross--Taylor string theory
interpretation in terms of branched covers, known for both the full theory
and for a chiral sector \cite%
{Gross:1992tu,Gross:1993hu,Cordes:1994fc,Donnelly:2016jet}. In turn, already
in the undeformed case, the interplay of all these aspects with the large $N$
phase transition is an interesting subject \cite{Crescimanno:1994eg}.
Another possibility could be to start with the generalized theory \cite{Douglas:1994pq,Ganor:1994bq}, which
also includes the effect of higher-order Casimir operators or perhaps
consider the theory with the presence of a topological $\theta $-term. These are well known aspects of the two-dimensional Yang--Mills
theory that would possibly be interesting to inspect from the novel point of
view of the $T\overline{T}$-deformation.

More recent developments, that nonetheless are already being investigated
for over a decade, involve the $q$-deformation of the $2d$ Yang--Mills
theory, which has been proven to enjoy a large number of relationships with
topological strings, Chern--Simons theories and supersymmetric gauge
theories in a number of dimensions (e.g. in the study of the $4d$
superconformal index \cite{Gadde:2011ik}). The partition function of $q$%
-deformed Yang--Mills theory on a Riemann surface $\Sigma _{h}$ is a
straightforward variation of the Migdal formula, which involves quantum
dimensions rather than ordinary dimensions of representations of the gauge
group \cite{Aganagic:2004js}:%
\begin{equation*}
\mathcal{Z}_{N,h}(q,p)=\sum_{R}\left( \dim _{q}R\right) ^{2-2h}q^{\frac{p}{2}%
C_{2}\left( R\right) }.
\end{equation*}%
This $q$-deformed gauge theory can be regarded as an analytic continuation
of Chern--Simons gauge theory on a Seifert fibration of degree $p$ over the
Riemann surface $\Sigma _{h}$. For genus $h=0$, the Seifert manifold is the
three-sphere $\cs^{3}$ for $p=1$, regarded as the Hopf fibration $\cs%
^{3}\rightarrow \Sigma _{0}=\cs^{2}$, and the lens space $L(p,1)=\cs^{3}/%
\Zint_{p}$ for $p>1$. Localization results have been proved also for the $q$%
-deformed case \cite{Beasley:2005vf,Kallen:2011ny} (see also \cite%
{Caporaso:2005ta} for explicit derivation in the case of the sphere). 

Notice that, naively, the deformation is \textquotedblleft
complementary\textquotedblright\ to the $T\overline{T}$ deformation, as it
involves only the dimensions part and not the Casimir part \cite%
{Szabo:2013vva}. The problem of phase transitions in the $q$-deformed theory
was studied, also following the classical works \cite%
{Douglas:1993iia,Gross:1994mr}, in \cite%
{Jafferis:2005jd,Arsiwalla:2005jb,Caporaso:2005ta}.

\acknowledgments

We thank Roberto Tateo, Stefano Negro and Riccardo Conti for correspondence
and comments. The work of MT was supported by the Funda\c{c}\~{a}o para a Ci%
\^{e}ncia e a Tecnologia (FCT) through its program Investigador FCT IF2014,
under contract IF/01767/2014. The work of LS was supported by the Funda\c{c}%
\~{a}o para a Ci\^{e}ncia e a Tecnologia (FCT) through the doctoral
scholarship SFRH/BD/129405/2017. The work is also supported by FCT Project
PTDC/MAT-PUR/30234/2017.

\begin{appendix}

\section{Instanton sectors and Poisson resummation}

\label{sec:InstantonPoissonResum}

This appendix is dedicated to a more accurate calculation of the instanton
contributions in the present model. We will follow the procedure of \cite%
{Minahan:1993tp} to evaluate the weights $w\left( \ell \right) $ in the $T%
\overline{T}$-deformed version of the theory \eqref{eq:ttZYM2}. The full
instanton expansion is obtained starting from the formula 
\begin{equation*}
\mathcal{Z}_{N}\left( A,\tau \right) =\sum_{\ell \in \Zint^{N}}Z_{\ell },
\end{equation*}%
where $Z_{\ell}$ is the Fourier transform of the sector corresponding to a given
representation, that is, 
\begin{equation}
Z_{\ell }=\int \prod_{i}dh_{i}e^{-2\pi \mathrm{i}\sum_{i}\ell
_{i}h_{i}}\prod_{i<j}\left( \frac{h_{i}-h_{j}}{i-j}\right) ^{2}e^{-\frac{A}{%
2N}\sum_{j=0}^{\infty }\left( \frac{\tau }{N^{3}}\right) ^{j}\left[
\sum_{i}\left( h_{i}-i+\frac{N}{2}\right) \left( h_{i}+i-\frac{N}{2}\right) %
\right] ^{j+1}}.
\end{equation}%
We can expand the contribution arising from the $T \overline{T}$-deformation: 
\begin{equation}
\begin{aligned} Z_{\ell } = \mathcal{C} \int & \prod_i dh_i e^{- 2 \pi \I
\sum_i \ell_i h_i } \prod_{i<j} \left( h_i - h_j \right)^2  e^{- \frac{A}{2N} \sum_{i} h_i ^2 }  \\ & \times
\left\{ \sum_{n=0}^{\infty} \frac{1}{n!} \left(- \frac{A}{2N} \right)^n \sum_{j=0} ^{\infty} c_j (n) \left( \frac{\tau}{N^3} \right)^j \left[  \sum_i h_i ^2 - \frac{N(N^2 -1)}{12} \right]^{j} \right\} ,
\end{aligned}  \label{eq:defFouriermode}
\end{equation}%
where $\mathcal{C}$ is an irrelevant overall factor and $\left\{ c_j (n) \right\}_j $ are the coefficients of the series expansion of $\left( \frac{x}{1- x} \right)^{n}$. Neglecting the shift in
the Casimir, and putting the focus on the last term, we obtain its Fourier
transform as: 
\begin{equation*}
\int dh_{i}e^{-2\pi \mathrm{i}\ell _{i}h_{i}-\frac{A}{2N}h_{i}^{2}}\left[ 
\frac{\tau }{N^{3}}h_{i}^{2}\right] ^{j}=\left( \frac{\tau }{%
N^{3}}\right) ^{j}\Gamma \left( \frac{1}{2} + {j}\right) {%
_{1}F_{1}}\left( \frac{1}{2}+ {j},\frac{1}{2},-\frac{N}{2A}\ell
_{i}^{2}\right) ,
\end{equation*}%
where ${_{1}F_{1}}$ is the confluent hypergeometric function.

The key observation to go further is that, inside the integral %
\eqref{eq:defFouriermode}, three ingredients appear: the Gaussian measure,
the Vandermonde determinant and a totally symmetric polynomial in the
variables $h_{i}^{2}$. Therefore, performing the integration with a single
Vandermonde determinant, which is a totally antisymmetric polynomial, one
obtains again the Vandermonde multiplying some totally symmetric polynomial
(or total symmetrisation of hypergeometric functions), up to overall
constant factors. The result is indeed: 
\begin{equation*}
\begin{aligned} & f_1 \left( \ell \right) := \int \prod_i dh_i e^{- 2 \pi \I \sum_i \ell_i h_i - \frac{A}{2N} \sum_i h_i ^2 }
\prod_{i<j} \left( h_i - h_j \right) \left\{ 
\sum_{n=0}^{\infty} \frac{1}{n!} \left(- \frac{A}{2N} \right)^n \sum_{j=0} ^{\infty} c_j (n) \left( \frac{\tau}{N^3} \right)^j \left[ 
\sum_i h_i ^2 \right]^{j} \right\} \\ & = \mathcal{C}^{\prime} \ \prod_{i<j}
\left( \ell_i - \ell_j \right) e^{- \frac{N}{2A} \sum_i \ell_i ^2 }
\sum_{n=0}^{\infty} \frac{1}{n!} \left(- \frac{A}{2N} \right)^n \sum_{j=0} ^{\infty} c_j (n) \left( \frac{\tau}{N^3}
\right)^{ j } \mathcal{S} \left( {_1 F _1} \left( \frac{1}{2} + \nu_i + j , \frac{1}{2} + \tilde{\nu}_i , - \frac{N}{2 A} \ell_i ^2 \right)
\right) , \end{aligned}
\end{equation*}%
where $\mathcal{S} (x) $ is a totally symmetric polynomial of order $N$ in $N$
variables, and $\nu _{i},\tilde{\nu}_{i}\in \mathbb{N}$. Therefore, the
arguments presented in \cite{Minahan:1993tp} hold also in this case. Taking
care of the polynomials that appear due to the $T\overline{T}$-deformation,
and one could retrieve the exact form of $Z_{\ell}$, by taking the convolution of two expressions as given above
\begin{equation*}
	Z_{\ell} = \left( f_1 \ast f_1 \right) \left( \ell \right) .
\end{equation*}

Although the discussion presented in this Appendix is qualitative, it
illustrates how the exact instanton contribution should in principle be
obtainable following standard methods. In particular, the shift in the quadratic Casimir can be reintroduced, and, using the binomial expansion, one can apply the calculations sketched here, taking care of the coefficients, to retrieve the exact contribution of each instanton sector.

\end{appendix}

\bibliographystyle{JHEP}
\bibliography{TTdef_biblio}

\providecommand{\href}[2]{#2}\begingroup\raggedright\begin{thebibliography}{10}

\bibitem{Zamolodchikov:2004ce}
A.~B. Zamolodchikov, \emph{{Expectation value of composite field T anti-T in
  two-dimensional quantum field theory}}, {\emph{{}} (2004) }
  [\href{https://arxiv.org/abs/hep-th/0401146}{{\ttfamily hep-th/0401146}}].

\bibitem{Smirnov:2016lqw}
F.~A. Smirnov and A.~B. Zamolodchikov, \emph{{On space of integrable quantum
  field theories}}, {\emph{Nucl. Phys.} {\bfseries B915} (2017) 363}
  [\href{https://arxiv.org/abs/1608.05499}{{\ttfamily 1608.05499}}].

\bibitem{Cavaglia:2016oda}
A.~Cavaglià, S.~Negro, I.~M. Szécsényi and R.~Tateo, \emph{{$T
  \bar{T}$-deformed 2D Quantum Field Theories}}, {\emph{JHEP} {\bfseries 10}
  (2016) 112} [\href{https://arxiv.org/abs/1608.05534}{{\ttfamily
  1608.05534}}].

\bibitem{Dubovsky:2017cnj}
S.~Dubovsky, V.~Gorbenko and M.~Mirbabayi, \emph{{Asymptotic fragility, near
  AdS$_{2}$ holography and $ T\overline{T} $}}, {\emph{JHEP} {\bfseries 09}
  (2017) 136} [\href{https://arxiv.org/abs/1706.06604}{{\ttfamily
  1706.06604}}].

\bibitem{Dubovsky:2018bmo}
S.~Dubovsky, V.~Gorbenko and G.~Hernández-Chifflet, \emph{{$T\bar{T}$
  Partition Function from Topological Gravity}}, {\emph{JHEP} {\bfseries 09}
  (2018) 158} [\href{https://arxiv.org/abs/1805.07386}{{\ttfamily
  1805.07386}}].

\bibitem{Giveon:2017nie}
A.~Giveon, N.~Itzhaki and D.~Kutasov, \emph{{$ \mathrm{T}\overline{\mathrm{T}}
  $ and LST}}, {\emph{JHEP} {\bfseries 07} (2017) 122}
  [\href{https://arxiv.org/abs/1701.05576}{{\ttfamily 1701.05576}}].

\bibitem{Giveon:2017myj}
A.~Giveon, N.~Itzhaki and D.~Kutasov, \emph{{A solvable irrelevant deformation
  of AdS$_{3}$/CFT$_{2}$}}, {\emph{JHEP} {\bfseries 12} (2017) 155}
  [\href{https://arxiv.org/abs/1707.05800}{{\ttfamily 1707.05800}}].

\bibitem{Datta:2018thy}
S.~Datta and Y.~Jiang, \emph{{$T\bar{T}$ deformed partition functions}},
  {\emph{JHEP} {\bfseries 08} (2018) 106}
  [\href{https://arxiv.org/abs/1806.07426}{{\ttfamily 1806.07426}}].

\bibitem{Aharony:2018bad}
O.~Aharony, S.~Datta, A.~Giveon, Y.~Jiang and D.~Kutasov, \emph{{Modular
  invariance and uniqueness of $T\bar{T}$ deformed CFT}}, {\emph{{}} (2018) }
  [\href{https://arxiv.org/abs/1808.02492}{{\ttfamily 1808.02492}}].

\bibitem{Cardy:2018sdv}
J.~Cardy, \emph{{The $T\overline T$ deformation of quantum field theory as
  random geometry}}, {\emph{{}} (2018) }
  [\href{https://arxiv.org/abs/1801.06895}{{\ttfamily 1801.06895}}].

\bibitem{Conti:2018tca}
R.~Conti, S.~Negro and R.~Tateo, \emph{{The $\textrm{T}\bar{\textrm{T}}$
  perturbation and its geometric interpretation}}, {\emph{{}} (2018) }
  [\href{https://arxiv.org/abs/1809.09593}{{\ttfamily 1809.09593}}].

\bibitem{Cardy:2018jho}
J.~Cardy, \emph{{$T\overline T$ deformations of non-Lorentz invariant field
  theories}}, {\emph{{}} (2018) }
  [\href{https://arxiv.org/abs/1809.07849}{{\ttfamily 1809.07849}}].

\bibitem{Conti:2018jho}
R.~Conti, L.~Iannella, S.~Negro and R.~Tateo, \emph{{Generalised Born-Infeld
  models, Lax operators and the $\textrm{T} \bar{\textrm{T}}$ perturbation}},
  {\emph{{}} (2018) } [\href{https://arxiv.org/abs/1806.11515}{{\ttfamily
  1806.11515}}].

\bibitem{Cordes:1994fc}
S.~Cordes, G.~W. Moore and S.~Ramgoolam, \emph{{Lectures on 2-d Yang-Mills
  theory, equivariant cohomology and topological field theories}}, {\emph{Nucl.
  Phys. Proc. Suppl.} {\bfseries 41} (1995) 184}
  [\href{https://arxiv.org/abs/hep-th/9411210}{{\ttfamily hep-th/9411210}}].

\bibitem{Douglas:1993iia}
M.~R. Douglas and V.~A. Kazakov, \emph{{Large N phase transition in continuum
  QCD in two-dimensions}}, {\emph{Phys. Lett.} {\bfseries B319} (1993) 219}
  [\href{https://arxiv.org/abs/hep-th/9305047}{{\ttfamily hep-th/9305047}}].

\bibitem{Gross:1994mr}
D.~J. Gross and A.~Matytsin, \emph{{Instanton induced large N phase transitions
  in two-dimensional and four-dimensional QCD}}, {\emph{Nucl. Phys.} {\bfseries
  B429} (1994) 50} [\href{https://arxiv.org/abs/hep-th/9404004}{{\ttfamily
  hep-th/9404004}}].

\bibitem{Migdal:1975zg}
A.~A. Migdal, \emph{{Recursion Equations in Gauge Theories}}, {\emph{Sov. Phys.
  JETP} {\bfseries 42} (1975) 413}.

\bibitem{Menotti:1981ry}
P.~Menotti and E.~Onofri, \emph{{The Action of SU($N$) Lattice Gauge Theory in
  Terms of the Heat Kernel on the Group Manifold}}, {\emph{Nucl. Phys.}
  {\bfseries B190} (1981) 288}.

\bibitem{Rusakov:1990rs}
B.~E. Rusakov, \emph{{Loop averages and partition functions in U(N) gauge
  theory on two-dimensional manifolds}}, {\emph{Mod. Phys. Lett.} {\bfseries
  A5} (1990) 693}.

\bibitem{Douglas:1993wy}
M.~R. Douglas, \emph{{Conformal field theory techniques in large N Yang-Mills
  theory}}, {\emph{{NATO Advanced Research Workshop on New Developments in
  String Theory, Conformal Models and Topological Field Theory Cargese,
  France}} (1993) } [\href{https://arxiv.org/abs/hep-th/9311130}{{\ttfamily
  hep-th/9311130}}].

\bibitem{Minahan:1993tp}
J.~A. Minahan and A.~P. Polychronakos, \emph{{Classical solutions for
  two-dimensional QCD on the sphere}}, {\emph{Nucl. Phys.} {\bfseries B422}
  (1994) 172} [\href{https://arxiv.org/abs/hep-th/9309119}{{\ttfamily
  hep-th/9309119}}].

\bibitem{Gross:1980he}
D.~J. Gross and E.~Witten, \emph{{Possible Third Order Phase Transition in the
  Large N Lattice Gauge Theory}}, {\emph{Phys. Rev.} {\bfseries D21} (1980)
  446}.

\bibitem{Wadia:1980cp}
S.~R. Wadia, \emph{{$N$ = Infinity Phase Transition in a Class of Exactly
  Soluble Model Lattice Gauge Theories}}, {\emph{Phys. Lett.} {\bfseries 93B}
  (1980) 403}.

\bibitem{Caselle:1993mq}
M.~Caselle, A.~D'Adda, L.~Magnea and S.~Panzeri, \emph{{Two-dimensional QCD on
  the sphere and on the cylinder}}, {\emph{{Proceedings, Summer School in
  High-energy physics and cosmology: Trieste, Italy}} (1993) 0245}
  [\href{https://arxiv.org/abs/hep-th/9309107}{{\ttfamily hep-th/9309107}}].

\bibitem{Gross:1992tu}
D.~J. Gross, \emph{{Two-dimensional QCD as a string theory}}, {\emph{Nucl.
  Phys.} {\bfseries B400} (1993) 161}
  [\href{https://arxiv.org/abs/hep-th/9212149}{{\ttfamily hep-th/9212149}}].

\bibitem{Gross:1993hu}
D.~J. Gross and W.~Taylor, \emph{{Two-dimensional QCD is a string theory}},
  {\emph{Nucl. Phys.} {\bfseries B400} (1993) 181}
  [\href{https://arxiv.org/abs/hep-th/9301068}{{\ttfamily hep-th/9301068}}].

\bibitem{Donnelly:2016jet}
W.~Donnelly and G.~Wong, \emph{{Entanglement branes in a two-dimensional string
  theory}}, {\emph{JHEP} {\bfseries 09} (2017) 097}
  [\href{https://arxiv.org/abs/1610.01719}{{\ttfamily 1610.01719}}].

\bibitem{Pipkin:1991}
A.~C. Pipkin, \emph{A Course on Integral Equations}, no.~9 in Texts in Applied
  Mathematics. Springer-Verlag, 1991.

\bibitem{Muskhe:1977}
N.~I. Muskhelishvili, \emph{Singular Integral Equations}. Noordhoff, 1977.

\bibitem{Jafferis:2005jd}
D.~Jafferis and J.~Marsano, \emph{{A DK phase transition in q-deformed
  Yang-Mills on S**2 and topological strings}}, {\emph{{}} (2005) }
  [\href{https://arxiv.org/abs/hep-th/0509004}{{\ttfamily hep-th/0509004}}].

\bibitem{Szabo:2013vva}
R.~J. Szabo and M.~Tierz, \emph{{q-deformations of two-dimensional Yang-Mills
  theory: Classification, categorification and refinement}}, {\emph{Nucl.
  Phys.} {\bfseries B876} (2013) 234}
  [\href{https://arxiv.org/abs/1305.1580}{{\ttfamily 1305.1580}}].

\bibitem{Witten:1991we}
E.~Witten, \emph{{On quantum gauge theories in two-dimensions}}, {\emph{Commun.
  Math. Phys.} {\bfseries 141} (1991) 153}.

\bibitem{Witten:1992xu}
E.~Witten, \emph{{Two-dimensional gauge theories revisited}}, {\emph{J. Geom.
  Phys.} {\bfseries 9} (1992) 303}
  [\href{https://arxiv.org/abs/hep-th/9204083}{{\ttfamily hep-th/9204083}}].

\bibitem{Blau:1993tv}
M.~Blau and G.~Thompson, \emph{{Derivation of the Verlinde formula from
  Chern-Simons theory and the G/G model}}, {\emph{Nucl. Phys.} {\bfseries B408}
  (1993) 345} [\href{https://arxiv.org/abs/hep-th/9305010}{{\ttfamily
  hep-th/9305010}}].

\bibitem{Blau:1993hj}
M.~Blau and G.~Thompson, \emph{{Lectures on 2-d gauge theories: Topological
  aspects and path integral techniques}}, {\emph{{Proceedings, Summer School in
  High-energy physics and cosmology: Trieste, Italy}} (1993) 0175}
  [\href{https://arxiv.org/abs/hep-th/9310144}{{\ttfamily hep-th/9310144}}].

\bibitem{Caporaso:2005ta}
N.~Caporaso, M.~Cirafici, L.~Griguolo, S.~Pasquetti, D.~Seminara and R.~J.
  Szabo, \emph{{Topological strings and large N phase transitions. I. Nonchiral
  expansion of q-deformed Yang-Mills theory}}, {\emph{JHEP} {\bfseries 01}
  (2006) 035} [\href{https://arxiv.org/abs/hep-th/0509041}{{\ttfamily
  hep-th/0509041}}].

\bibitem{Giombi:2009ds}
S.~Giombi and V.~Pestun, \emph{{Correlators of local operators and 1/8 BPS
  Wilson loops on S**2 from 2d YM and matrix models}}, {\emph{JHEP} {\bfseries
  10} (2010) 033} [\href{https://arxiv.org/abs/0906.1572}{{\ttfamily
  0906.1572}}].

\bibitem{Giombi:2009ek}
S.~Giombi and V.~Pestun, \emph{{The 1/2 BPS 't Hooft loops in N=4 SYM as
  instantons in 2d Yang-Mills}}, {\emph{J. Phys.} {\bfseries A46} (2013)
  095402} [\href{https://arxiv.org/abs/0909.4272}{{\ttfamily 0909.4272}}].

\bibitem{Bassetto:1998sr}
A.~Bassetto and L.~Griguolo, \emph{{Two-dimensional QCD, instanton
  contributions and the perturbative Wu-Mandelstam-Leibbrandt prescription}},
  {\emph{Phys. Lett.} {\bfseries B443} (1998) 325}
  [\href{https://arxiv.org/abs/hep-th/9806037}{{\ttfamily hep-th/9806037}}].

\bibitem{Drukker:2000rr}
N.~Drukker and D.~J. Gross, \emph{{An Exact prediction of N=4 SUSYM theory for
  string theory}}, {\emph{J. Math. Phys.} {\bfseries 42} (2001) 2896}
  [\href{https://arxiv.org/abs/hep-th/0010274}{{\ttfamily hep-th/0010274}}].

\bibitem{Durhuus:1980nb}
B.~Durhuus and P.~Olesen, \emph{{The Spectral Density for Two-dimensional
  Continuum {QCD}}}, {\emph{Nucl. Phys.} {\bfseries B184} (1981) 461}.

\bibitem{Blaizot:2008nc}
J.-P. Blaizot and M.~A. Nowak, \emph{{Large N(c) confinement and turbulence}},
  {\emph{Phys. Rev. Lett.} {\bfseries 101} (2008) 102001}
  [\href{https://arxiv.org/abs/0801.1859}{{\ttfamily 0801.1859}}].

\bibitem{Neuberger:2008ti}
H.~Neuberger, \emph{{Complex Burgers' equation in 2D SU(N) YM}}, {\emph{Phys.
  Lett.} {\bfseries B670} (2008) 235}
  [\href{https://arxiv.org/abs/0809.1238}{{\ttfamily 0809.1238}}].

\bibitem{Crescimanno:1994eg}
M.~J. Crescimanno and W.~Taylor, \emph{{Large N phases of chiral QCD in
  two-dimensions}}, {\emph{Nucl. Phys.} {\bfseries B437} (1995) 3}
  [\href{https://arxiv.org/abs/hep-th/9408115}{{\ttfamily hep-th/9408115}}].

\bibitem{Douglas:1994pq}
M.~R. Douglas, K.~Li and M.~Staudacher, \emph{{Generalized two-dimensional
  QCD}}, {\emph{Nucl. Phys.} {\bfseries B420} (1994) 118}
  [\href{https://arxiv.org/abs/hep-th/9401062}{{\ttfamily hep-th/9401062}}].

\bibitem{Ganor:1994bq}
O.~Ganor, J.~Sonnenschein and S.~Yankielowicz, \emph{{The String theory
  approach to generalized 2-D Yang-Mills theory}}, {\emph{Nucl. Phys.}
  {\bfseries B434} (1995) 139}
  [\href{https://arxiv.org/abs/hep-th/9407114}{{\ttfamily hep-th/9407114}}].

\bibitem{Gadde:2011ik}
A.~Gadde, L.~Rastelli, S.~S. Razamat and W.~Yan, \emph{{The 4d Superconformal
  Index from q-deformed 2d Yang-Mills}}, {\emph{Phys. Rev. Lett.} {\bfseries
  106} (2011) 241602} [\href{https://arxiv.org/abs/1104.3850}{{\ttfamily
  1104.3850}}].

\bibitem{Aganagic:2004js}
M.~Aganagic, H.~Ooguri, N.~Saulina and C.~Vafa, \emph{{Black holes, q-deformed
  2d Yang-Mills, and non-perturbative topological strings}}, {\emph{Nucl.
  Phys.} {\bfseries B715} (2005) 304}
  [\href{https://arxiv.org/abs/hep-th/0411280}{{\ttfamily hep-th/0411280}}].

\bibitem{Beasley:2005vf}
C.~Beasley and E.~Witten, \emph{{Non-Abelian localization for Chern-Simons
  theory}}, {\emph{J. Diff. Geom.} {\bfseries 70} (2005) 183}
  [\href{https://arxiv.org/abs/hep-th/0503126}{{\ttfamily hep-th/0503126}}].

\bibitem{Kallen:2011ny}
J.~Kallen, \emph{{Cohomological localization of Chern-Simons theory}},
  {\emph{JHEP} {\bfseries 08} (2011) 008}
  [\href{https://arxiv.org/abs/1104.5353}{{\ttfamily 1104.5353}}].

\bibitem{Arsiwalla:2005jb}
X.~Arsiwalla, R.~Boels, M.~Marino and A.~Sinkovics, \emph{{Phase transitions in
  q-deformed 2-D Yang-Mills theory and topological strings}}, {\emph{Phys.
  Rev.} {\bfseries D73} (2006) 026005}
  [\href{https://arxiv.org/abs/hep-th/0509002}{{\ttfamily hep-th/0509002}}].

\end{thebibliography}\endgroup

\end{document}